# Advanced Power Transmission of the Future

Mario Rabinowitz
Armor Research, 715 Lakemead Way, Redwood City, CA 94062-3922
Mario715@earthlink.net

**Abstract**

Electric power is a vital ingredient of modern society.  This article is written to provide an insight into the physics and engineering that go into the transmission of electric power and its potential modernization. Topics covered will be Transmission and Distribution, Comparing Overhead and Underground Delivery, Pros and Cons of Underground Delivery, Superconducting Transmission, Cryorisistive Delivery, Hyperconductivity, and Metal-Plated Graphite Fibers.

**Perspective**

Modernization of electric power  systems in the next two decades will be key to improving the standard of living of future generations, as well as decreasing the vulnerability of civilization to the onslaughts of terrorism.  There is a potential for smart cables that can aid in the location of a fault, and possibly (by no means out of the question) aid in the detection and location of incipient faults so they may be repaired in the course of routine maintenance rather than after they have caused expensive failures. Substantially more compact transformers are possible if we are willing to do the necessary R & D to develop them.   A consequent advantage may be lower losses. New developments using combinations of additives show great promise in solving the transformer oil flow electrification problem.  We will be able to supply custom power for manufacturers that require and value it.  Superconducting fault current limiters that not only protect the system but that also ease the burden on circuit breakers are a definite possibility.  Greatly improved restoration preparedness will be widely adapted. Excavation for underground lines will reach new heights of sophistication.

Hyperconducting materials may greatly reduce power losses and increase power density.  Proper precautions have been developed for their manufacture, as is the case for highly toxic materials in semiconductor production. Even if they will be proven to be safe in the field, there are legitimate questions of public acceptance to which utilities will have to be increasingly responsive. Heightened public awareness of electromagnetic field issues will also continue to be addressed. Distribution automation will receive a favorable public reaction, and enhance power delivery.

Power systems of the future should enable utilities to:
•be more competitive with their overall strategies, •provide better service, •better manage their assets, •extend equipment life, •improve diagnostics, •develop reliability-centered maintenance.

In looking ahead at the next two decades, we will consider various options that are available, point out the less likely, and highlight those that have the potential of achieving major improvements. Although the next 20 years is a little soon for dispersed generation to have a significant impact, some of the implications of dispersed generation are considered. The Preview of Contents that follows is indicative of the broad range of topics that will be covered in Parts I to IV.

**Introduction**

During the 1960's, the generating capacity of the U. S. electric power industry almost doubled, growing from 175 to 325 GW (gigawatt = $10^9$ W). In 1974, it stood at 474 GW. By 1980, it had reached 600 GW. By the end of 1993, it reached 700 GW. It is expected that over 210 GW of new capacity will be required by 2010, bringing the U. S. capacity for the first time into the TW (terawatt = $10^{12}$ W) range. According to Electrical World, less than 20% of this needed new electric capacity is under construction.[1]

The current relatively modest growth in demand of 2% per year could dramatically increase should electricity consumption grow from the present 30% to an expected 50% of all forms of energy consumption by the year 2030. Increased population, with a corresponding increase in population density, will contribute to this greater rate, due significantly to the lower cost, versatility, convenience and safety of electric energy. General increase in the standard of living will also be an important driver in the growth in use of electrical power.

Increased electricity consumption has generally followed the growth of the Gross Domestic Product, even when total energy consumption remains flat. With the pronounced movement of the industry toward deregulation, intensified competition and retail wheeling, these increased demands need to be accurately forecast. Regulation with respect to rights-of-way and large capital outlays further adds to the need to anticipate increased power delivery well in advance. Since these and many other factors apply to both transmission and distribution systems, a strong distinction will not be made between them at this time, and the more general term delivery will often be used.

Prior to the worldwide energy crisis of 1974, electricity consumption in both the United States and in western Europe nearly doubled every 10 years, at an annual growth rate of about 7%. For a number of years following 1974, many combined factors reduced this growth rate to 3% annually in the United States. Currently, the average growth rate in domestic electric demand hovers around 2%. This relatively modest growth in demand could dramatically increase should electricity consumption grow from the present 30% to an expected 50% of all forms of energy consumption by the year 2030. Increased population, with a corresponding increase in population density, will contribute to this greater rate, due significantly to the lower cost, versatility, convenience and safety of electric energy. General increase in the standard of living will also be an important driver in the growth in use of electrical power.

**Transmission and Distribution**

Although traditionally high capital costs have been acceptable for transmission lines (> 35 kV), they have not been for distribution lines (≤ 35 kV) with capital costs as low as $1 to $2/foot ($5000 to $10,000/mi). Thus many new power delivery technologies which are capital intensive are usually eliminated from consideration as potential distribution systems.  However, with increasing demand for greater reliability, decreased power losses, lower operating and maintenance costs, awareness of possible biological effects of electromagnetic fields, and greater longevity of the lines, we should also look at emerging technologies that may have application to distribution despite their initial high costs.

We have learned a lesson from all of the questionable distribution cable that was put in service 15 to 20 years ago that initially low capital costs may be misleading, since replacement costs can be quite high.  For those utilities that must consider the short term, lower initial capital cost may dominate their perspective.  However, for those utilities that have the foresight and can afford to pursue the long-term, considerations of reliability, lifetime, convenience, and the entire genre of maintenance, operations, and installation as well as initial capital costs will dominate their implementation of a truly modern power delivery system.

Since the beginning of the utility industry, insulation has been the flesh and a simple good conductor such as copper or aluminum has been the backbone of power delivery.   Sodium has been used successfully on a small scale, but has not received wide acceptance because it burns when exposed to air.  Desirable conductor properties include low density, reasonably high conductivity, low cost, and chemical stability.  In simple terms as a figure of merit, it is desirable that the conductivity divided by the density be as large as possible.  This figure can be further divided by cost for an economic comparison.  In this framework, a material like sodium would look good if it didn't burn, since it is about 1/9 as dense with a conductivity about 1/3 that of copper, giving it a figure of merit 3 times better than copper.  Since air is the main dielectric for overhead lines, tensile strength is also an important factor in this case because there is no mechanical support from a solid dielectric.  Here conducting polymers might look good, were it not for their chemical instability.  This will be discussed in a later issue.

Power delivery has and will play an increasingly important role in the utility industry. In the first half of this century, growth in transmission line capacity was directly proportional to the increase in turbogenerator and power plant sizes. Due to economies of scale and increased demand, turbo-generators increased in ratings from about 1 MVA and 10 kV terminal voltage in the early 1900's to the present 1500 MVA and 25 kV with few substantial changes in technology. As the generating units and power plants increased in size, so too did the capacity of the transmission lines. To reduce losses, as more power was carried by a given line, it became desirable to deliver the power at ever higher voltages. Thus voltage ratings of transmission lines in the U. S. increased from 10 kV to 765 kV. This required high capacity step-up transformers to connect the generators to the transmission grid, and high capacity step-down transformers at sub-transmission and distribution stations. In less than a century, transmission line capabilities increased from 1 MVA to over 1500 MVA. This level represents about the largest amount of power that a utility is willing to carry on one line because of reliability concerns. The risk of losing this much power due to line failure, together with other contingencies, is too great a liability for most utilities.

**Comparing Overhead and Underground Delivery**

Excluding rights-of-way costs, the installation (and presumably maintenance costs also) of overhead lines has been, and probably will continue to be, less than that of underground lines. Because power line corridors come at a premium due to the burdensome cost of rights-of-way, overhead lines can be as expensive as underground lines in densely populated areas. However, overhead lines are not a form of high-density power transmission. Any kind of high-density power transmission, especially underground lines, requires expensive cooling. An overhead line is effectively cooled by the large air space surrounding it, which is required for dielectric strength, and to keep the electric field within code at ground level. However, despite the lower cost of most overhead lines, it is likely that a decreasing proportion of power will be transmitted overhead because of ecological, practical, and aesthetic considerations which are reflected in the difficulty of obtaining new rights-of-way.  So, except for upgrading existing overhead lines, much of the new delivery capacity in the future will be underground -- possibly sharing the same corridors with overhead lines.

Due in part to their simplicity, lower cost, and ease of repair, overhead lines have been the dominant form of delivery from the earliest days to the present.  Nevertheless, underground lines have seen increasing application in many areas of the U. S. due to their greater reliability.  However, such reliability comes at substantially greater installed cost per mile.  Though faults may occur more frequently on overhead lines, they are more easily detected, located and repaired.  It may be difficult to make a meaningful comparison of fault frequency, fault duration, cost of fault repair, and operating costs between overhead and underground delivery because the comparison needs to be site specific.  Even if we normalize with respect to faults/mile, faults/MVA-mile, or repair costs/mile, it may turn out to be like comparing apples and oranges because of differences in terrain and weather.

Overhead faults last from seconds to days depending on the cause, with the longer duration caused by extremes of weather such as heavy ice loading, tornadoes, etc. Overhead outage rates for 138 kV and lower voltage lines are about 4 - 6 per 100 miles per year. The design target for higher voltage overhead lines is not to exceed one outage per 100 miles per year for lightning, and one outage per 1000 switching operations for switching overvoltage operations.  In both cases, an outage is a breaker operation, not total line disconnect.  There is a paucity of information available on outage rates, duration, and costs in the open literature for both overhead and underground.  There appears to be no publicly available reliable source of this data for either transmission or distribution, except that kept by individual utilities and provided through their good auspices.

According to Tom Rodenbaugh [2], overhead faults probably occur ~ 100 times more often than underground. This is usually due to wind loading, lightning strikes, and tracking on dirty insulators. Salt spray also is a major cause of faults along the coastline. Most overhead line faults last only a second or two. Most are intercepted by fault limiters and relays, and easily cleared. Underground transmission circuits have a fault about once per 1000 miles per year. This would be for 138 kV and above, whereas 46 kV, 69 kV, and 115 kV have more faults because they are either directly buried, or are placed in common duct banks with lower voltage distribution cables. Their frequency of faults is roughly double at ~ one per 500 miles per year. As a comparison, distribution cables fail about once per 100 miles per year.

Rodenbaugh observes that underground faults have a much longer average repair time of about a week; and that the cost associated with repair of an underground circuit can be much greater than for an overhead line. This is due to location difficulties, excavation if needed, and the fact that most splicing can only be done by specialized repair crews, who may not be immediately available. He estimates that a single phase fault in a pipe-type transmission circuit would cost $15,000 to $50,000 depending on the severity.

**Pros and Cons of Underground Delivery**

Although conventional self-cooled or force-cooled underground high-density power delivery does not always have the ecological and aesthetic drawbacks of overhead transmission, it can have other disadvantages. It is expensive to manufacture, install, and operate; and its high capital and trenching costs results mainly from the technical complexity of high-voltage insulation technology and its associated cooling requirements. (Leakage of cooling oil can be a serious ecological problem.) The high operating costs result from relatively high charging currents associated with the high voltage and high capacitance characteristics of underground lines, and cooling (refrigeration) inefficiency in the case of forced-cooled lines. Excavation of large trenches, special accessory equipment, and the introduction of high-thermal-conductivity materials to prevent thermal runaway often make the installation cost of underground transmission as high as the cost of the cable alone. The significantly lower power density of underground distribution cable greatly reduces the installation costs relative to transmission cable.

As available corridors become saturated and power dissipation increases as fast or faster than the increase in capacity, more attention will need to be given to the thermal conductivity of the backfill. Presently, a weak concrete slurry mix somewhat alleviates this problem. A special universal light weight hygroscopic backfill, that can easily be shipped anywhere, is worth pursuing. EPRI helped to develop a slack wax which can stabilize the thermal conductivity of soil. During dry hot periods, the moisture between backfill particles evaporates away leaving air gaps which result in a high thermal resistivity. The wax fills in these pores providing a thermal bridge between grains. Slack wax is an inexpensive by-product of oil refining that is stable in the ground, and is readily available in all parts of the U.S. It can be added to backfill in emulsified form, or by heating.

Because an underground line incurs more kinds of losses than an overhead line and their impact is greater, it may require as much as five times the cross-sectional area of conductor as an overhead line to carry the same power. Thus by an inverse convolution of cause and effect, many underground lines have total losses lower than overhead lines. A typical loss factor of an ac overhead line is 4.4% per 100 miles at 345 kV. Comparable ac underground lines have losses of about 3.5% per 100 miles. However at 500 kV, the ac overhead losses are down to 2.5% per 100 miles. The losses for 400 kV dc overhead are lower than 1% per 100 miles .

**Superconducting Transmission**

Since a good conductor is a key element of power delivery, let us look at the best conductors known, superconductors. Until late 1986, superconductivity was considered a very low temperature phenomenon near absolute zero (0 K= -273.2ºC = -459.7ºF). The accepted highest transition temperature, $T_c$ (temperature at which a material becomes a superconductor, also called the critical temperature) was found for $Nb_3Ge$ at 23.2 K in 1973, where it stayed for nearly 13 years. There were many reports of higher temperature superconductivity from 1973 to 1986, but these findings could not be replicated. Thus $T_c$ only increased by 19 K, from 4.2 K in 1911, in 75 years. This suggests a projected increase of about 25 K per century. From a misleading linear extrapolation, one might have expected the present, easily reproduced $T_c$ = 125 K for TlCaBaCuO to be achieved in about another four centuries. Instead, to the scientific world's amazement, a gain of about 70 K occurred in a matter of months between late 1986 and early 1987. Such revolutions, though they do not occur frequently, do occur regularly in science.

Superconductors only have infinite conductivity for dc below a critical current density, $J_c$, at which point resistivity sets in. High temperature superconductors generally are very poor conductors in the normal state, and have poor thermal conductivity in both states. For ac, there is a power loss in superconductors for all values of current density. In a coaxial transmission line, with the conductor exposed to only its relatively small tangential self-magnetic field, the power loss can be acceptably small. Interestingly, this hysteretic power loss is inversely proportional to $J_c$. So for both dc and ac, it is important to have as high a value of $J_c$ as possible. However, not as much progress in bulk superconductors has been achieved with critical current density, $J_c$, as with $T_c$. What is not well known is that superconductors are rather poor conductors in the normal state, which adds to the difficulty of incorporating them into high power applications.

Let us explore whether the near-term status of superconducting technology is relevant to the power delivery needs of the electric utility industry. Low-temperature superconducting lines (LTSL) have already proven to be technically feasible. LTSL's economic viability, however, remains in doubt. For a large surface-to-volume application like a delivery line, refrigeration efficiency is crucial. The ~ 700W/W of refrigeration needed for LTSL at ~ 4K operation tends to make it economically uncompetitive. The higher the $T_c$ of the superconductor, the higher the operating temperature of the cable, with obvious reductions in refrigeration costs. A fringe benefit of higher temperature operation is that the heat capacity of both superconductor and cryogen scales approximately as the cube of the absolute temperature. Thus with new materials, high-temperature superconducting lines (HTSL) operating at ~ 77 K with only ~ 10W/W of needed refrigeration may offer an attractive alternative to conventional underground delivery.

$Bi_{1.6}Pb_{0.4}Sr_2Ca_2Cu_3O_{10}$ and its variations (commonly referred to as BSCCO) with a $T_c$ of about 107 K are being considered HTSL. Because of the low thermal conductivity and low normal state electrical conductivity of high-temperature superconductors (HTSC), superconducting filaments are embedded in a silver matrix (sheath) where there is about 4 times as much silver as superconductor by volume. For utilities, power delivery applications may have the least number of technical obstacles to overcome in utilizing HTSC. (Although superconducting lines are inherently high current and low voltage, they have not customarily been considered for distribution because of their intrinsically high capital cost.) Nevertheless, technical problems remain formidable, and even a promising delivery application may not become a commercial reality within the next 20 years.

For a new material to be suitable for an ac HTSL, key issues that need to be addressed are brittleness, ac power loss, and critical current density, $J_c$. So far, all the high temperature superconductors are quite brittle. Even if a sufficiently high (> $10^5$ A/cm$^2$ overall) $J_c$ can be attained in cable form, the power losses may still be excessively high for non-coaxial designs being pursued for retrofit applications. (Bear in mind that for low temperature superconductors, $J_c$ > $10^6$ A/cm$^2$ in cable form, and the designs were coaxial.) Power losses are exacerbated if the magnetic field seen by the conductor is not tangential. In a coaxial design, the only magnetic field that a conductor sees is its azimuthal self-field which is tangential to the conductor. Thus for a coaxial cable, the power dissipation for the three phases is just three times that of one phase, because there is no coupling between the phases.

In a non-coaxial design, because of normal components of magnetic field and coupling between the phases, the power dissipation for the three phases is much more than three times that of one phase. Superconducting filaments in a helical tape structure add to the losses. Even if the design were coaxial, the filaments produce normal (perpendicular) components of magnetic field on each other as well as large eddy current losses in the normal silver matrix because the filaments do not shield the silver from the field as would a superconducting tape where the superconductor shields the normal conductor. The helical conductor geometry produces greater losses than would a monolithic cylinder. The presence of many filaments (rather than just one) has a small advantage as well as a large disadvantage associated with it. The advantage is that if one or more filaments break, the current can be shunted through a short distance of the normal silver to the other broken end, and/or other filaments. However, this bridging function is outweighed by considerably higher eddy current losses. When there are two or more filaments, the induced eddy current is much higher as the current goes up one filament and down the other, bridging across through the silver. The eddy current loss would be much smaller in the silver without the many superconducting filaments, as the current would then be resistance limited. The multifilamentary design is good for dc magnets, but not for ac. Because of this far-from-optimal design, the original hope that power loss for a superconducting cable will be less than for a normal cable will be difficult to achieve, especially while attempting to carry significantly more power than a normal cable.

Despite the formidable handicaps for the superconducting non-coaxial design that is somewhat like a three-in-one oil-filled pipe-type cable, it does have a dielectric advantage. The insulation can be at ambient temperature. This simplifies the terminations (potheads), and makes it easier to make splices. However, the disadvantage of much greater power losses may well outweigh this advantage, and limit the applications of this design to retrofitting the insides of existing pipe-type cables. Pulling the three fragile superconducting phases through the tight space of the existing pipe will be a difficult matter, particularly around bends.

Substantial R & D expenditures on large prototypes have been justified with the argument that this is necessary to determine compatibility with existing systems. This is certainly necessary when the technology is proven. However, it may be premature to do so when it is not even clear that the power losses will be less than for conventional systems.

If reduction in losses were the only consideration, this could be done with conventional lines by going to higher voltages (the traditional approach), increasing the amount of conductor in a given line, or increasing the number of lines. However, power density, reliability, and capital cost are foremost considerations. Thus in spite of its greater complexity, a superconducting line needs to be equally reliable, be no more expensive than its conventional counterparts at the same power rating, have higher power density, and hopefully have lower power loss. For present materials and designs, a superconducting line will likely have higher power density, but may not have lower losses than the much cheaper and simpler cryoresistive cable which will be discussed next.

**Cryoresistive Delivery**

**Conventional cryoresistive**

Electrical conductivity is proportional to the electron mean free path in a conductor, which increases as the lattice vibrational (phonon) scattering of the electrons decreases with decreasing temperature. In going from 300 K to about 77 K, the electron mean free path (and hence the conductivity) increases by approximately a factor of 10 for almost all materials, fairly independent of impurities and other lattice defects because phonon scattering dominates for them. (There is one exception, whose exceptional increase in conductivity will be discussed in the next section.) Liquid nitrogen ($LN_2$), which boils at 77 K at atmospheric pressure, is the coolant of choice for both HTSL and cryoresistive delivery.

The increase in conductivity by a factor of 10 permits a tenfold increase in power carried, but not without increase in power loss. In addition to resistive and dielectric power losses, there are also heat leak losses associated with the cryogen. The reason for the increased overall power loss (despite the increased conductivity) relates to power loss in the refrigeration system, which requires about 10W of refrigerator power for every watt dissipated at 77K.

All cryoresistive projects to date have been ac at liquid nitrogen temperature. The most active work has been in Japan. In the United States, the General Electric Company and Underground Power Corp. have worked in this area, but these projects have been abandoned due to technical and economic barriers. A similar fate seems to have befallen such projects worldwide. Aluminum (Al) was the conductor of choice, although copper (Cu) was also considered. Both would give an increase in conductivity of only a factor ~ 10 at 77 K. Let's take a look at two novel systems that could do much better than this, but that were not considered when the cryoresistive research was being done.

**Hyperconductivity**

One element stands out as having the highest gain in conductivity by a wide margin than all the rest when cooled to $LN_2$ temperature. This is beryllium (Be), which has a conductivity at room temperature that is comparable to Al, but which increases its conductivity by a factor of ~100 at 77 K. So at 77 K, Be has about 10 times the conductivity of Al or Cu. This was noted by Rabinowitz for cryogenic power applications in 1977.[3] In 1990, Mueller et al [4] argued that "beryllium should be considered for some conduction applications [at 77 K], despite its well known toxicity problems." They point out that Be is only hazardous in the form of fine airborne particulates, requiring careful control during the manufacturing process, but that Be would be relatively safe as a finished product in electrical applications.

The low density of Be of 1.8 gm/cm$^3$ is another property that makes it outstanding for conductors. This is less than the density of Al (2.7 gm/cm$^3$), and 5 times less than that of Cu (8.9 gm/cm$^3$). It is less than twice the density of sodium (0.97 gm/cm$^3$), and does not burn in air. Beryllium's conductivity even at room temperature is almost twice that of sodium. Therefore at 77 K, Be's conductivity/gm is so superior to other metallic conductors that it should seriously be considered for cryoresistive delivery and other cryogenic applications, bearing in mind its toxicity and expense. The toxicity is possibly only a problem in fabrication, where careful control in compliance with some of the earliest environmental laws seems to have overcome this problem. The cost of Be may well go down if a sufficiently large market develops. A serious look at the potential of hyperconducting Be has not yet been taken.

**Metal plated graphite fibers**

In going from 300 K to below 4.2 K (the boiling point of liquid helium at 1 atmosphere), the conductivity of most metals can increase by as much as a factor of 100 to 10,000, depending on purity and degree of lattice perfection. An increase in conductivity by a factor of 1000 or more could result in a corresponding increase in deliverable power for dc transmission. This increase would not be as readily achieved in the ac case because the skin depth is inversely proportional to the square root of the conductivity. Thus, the effective gain increases only as the square root of the resistivity ratio, unless very thin transposed wires are used. Another problem is that the degree of lattice perfection required to yield a large gain in low-temperature conductivity results in a material that has extremely low tensile and shear strength. In addition to the greater expense of making it, there would be the additional expense of mechanical reinforcement. Yet the very large gains in conductivity may warrant low temperature operation in certain circumstances.

High purity Al is readily available. Al and other high purity metals such as Cu have a very high increase in conductivity of a factor of 10,000 at < 4 K. And there is a solution for the very low tensile strength problem. Techniques have been developed for plating continuous thin coatings of copper, brass, silver, gold, and nickel onto graphite fibers. Other metals are also possible, but it is not clear whether aluminum and beryllium are among them. It is reasonable to assume that they could be. Graphite fibers are known for their light weight and exceptional tensile strength. The coated fibers with a diameter of 8 microns, metal coating of about 0.5 microns thick, density of 2.5 - 3 gm/cm$^3$ have a tensile strength up to 450,000 psi. This is a reduction of only 8 - 10 % from the original graphite fiber. By comparison, ordinary steel has a tensile strength between 40,000 and 330,000 psi. The best steel wire has a tensile strength up to 460,000 psi. So now the possibility exists for using the coated fibers directly, or embedding the fibers in bulk metallic conductor to make a high tensile strength matrix. One may think of using high purity reinforced metals like Al as the normal stabilizer in low temperature superconductivity applications, and as the conductor in cryoresistive applications. For low temperature ac, one might use transposed electroplated graphite fibers coated with a thin dielectric coating over the conductor to keep the inductance low.

If high power density is the main need, then operation well below 77 K in retrofit situations (such as retrofitting oil-cooled 3-in-1 pipe-type cables) may be an application for this technology. The potential for metal plated graphite fibers exists not only for cryoresistive applications, but also at ambient temperature in overhead lines where light weight and high tensile strength are important factors. Graphite fibers have been used successfully in the automotive and other industries as a lightweight means of increasing mechanical strength.

As exciting as the possibility of graphite fiber reinforced conductors may be, there is an even more exciting possibility. The recently discovered carbon nanotubes which have tremendous tensile strength may one day replace the much thicker graphite fibers that are now used to make ultra-light bicycles, tennis rackets, rocket motor casings, and perhaps will be used to reinforce electrical conductors. Nanotubes are named for their nanometer-size diameters ($10^{-9}$ m = 10 Å), with walls as thin as a single atom. The strong sorption capability of nanotubes related in part to their capillary action may make them easy to incorporate into electrical conductors to greatly increase their tensile strength. Nanotubes might also serve as casting molds for molten metal, which upon solidification would result in the finest wires possible. Such wires could be used in microscopic circuits. The commercial cost of nanotubes is expected to be modest, as they can be produced by the pound.

**References**

1. Electrical World, **207**, 17, Nov. 1993.
2. Private communication from Thomas J. Rodenbaugh of EPRI.
3. M. Rabinowitz, Cryogenics, **17**, 319 (1977).
4. F.M. Mueller et al (15 authors), Applied Physics Letters, **57**, 240 (1990).